\documentclass[twocolumn, showpacs, amsmath,amssymb]{revtex4} %envcountsame, fleqn
\usepackage{epsfig, amssymb}
\begin{document}
%\preprint{xxx}

\title{Local invariants of stabilizer codes}
%\\(Paper in progress, week report 19/03/2004)

\author{Maarten Van den Nest}
\email{maarten.vandennest@esat.kuleuven.ac.be}
\author{Jeroen Dehaene}
\author{Bart De Moor}
\affiliation{Katholieke Universiteit Leuven, ESAT-SCD, Belgium.}%Lines break automatically or can be forced with \\
\date{\today}
\begin{abstract}
In [Phys. Rev. A 58, 1833 (1998)] a family of polynomial invariants which separate the orbits of
multi-qubit density operators $\rho$ under the action of the local unitary group was presented. We
consider this family of invariants for the class of those $\rho$ which are the projection operators
describing stabilizer codes and give a complete translation of these invariants into the binary
framework in which stabilizer codes are usually described. Such an investigation of local
invariants of quantum codes is of natural importance in quantum coding theory, since locally
equivalent codes have the same error-correcting capabilities and local invariants are powerful
tools to explore their structure. Moreover, the present result is relevant in the context of
multipartite entanglement and the development of the measurement-based model of quantum computation
known as the one-way quantum computer.
\end{abstract}

\maketitle

\section{Introduction}
The theory of quantum error-correcting codes constitutes a vital ingredient in the realization of
quantum computing, as these codes protect the vulnerable information stored in a quantum computer
from the destructive effects of decoherence. The most widely known class of error-correcting codes
is that of the stabilizer codes, studied extensively in e.g. \cite{Gott,codeGF4, QCQI}. An
$n$-qubit stabilizer code is defined as a simultaneous eigenspace of a set of commuting observables
in the Pauli group, where the latter consists of all $n$-fold tensor products of the Pauli matrices
and the identity. Equivalently, a code is described by the projector operator on this eigenspace.
In the characterization of the the error-correcting capabilities of quantum codes, two equivalence
relations arise naturally on the set of corresponding projectors: two $n$-qubit quantum codes
described by projectors $\rho$ and $\rho'$ are called \emph{globally equivalent}, or just
\emph{equivalent}, if there exists a local unitary operator $U\in U(2)^{\otimes n}$ such that
$U\rho U^{\dagger}$ is equal to $\rho'$ modulo a permutation of the $n$ qubits. If $U\rho
U^{\dagger}=\rho'$, without any additional permutation, the codes are called \emph{locally
equivalent}. As globally equivalent codes have exactly the same error-correcting capabilities and
vice versa, global equivalence is in fact the true equivalence of quantum codes. However, the
structure of local equivalence is more transparent and insight in this matter already provides a
lot of information about the structure of quantum codes. Therefore, much of the relevant literature
tackles local equivalence and we will do the same in the following.

This paper is concerned with the characterization of the local equivalence class of a stabilizer
code $\rho$ by means of local \emph{invariants}. These are complex functions $F(\rho)$ which remain
invariant under the action of all local unitary transformations, i.e., \[F(\rho)=F(U\rho
U^{\dagger})\] for every $U\in U(2)^{\otimes n}$. In studying invariants, the general goal is to
look for a minimal set of invariants which characterizes the local equivalence class of any given
code. To obtain such a minimal complete set, it is well known \cite{pol_inv_suff} that it is
sufficient to consider functions $F$ which are polynomials in the entries of $\rho$.  These
polynomial invariants form an algebra over $\Bbb{C}$, as linear combinations and products of
invariants remain invariants. Interestingly, the invariant algebra of $U(2)^{\otimes n}$ is
finitely generated \cite{finit_gen} and therefore the existence of a \emph{finite} complete set of
polynomial invariants is guaranteed. Although the problem of pinpointing such a finite set is to
date unanswered, progress has been made in the past years in constructing complete though infinite
families of invariants. A natural approach is to consider homogeneous invariants, since any
invariant can be written as a sum of its homogeneous components, each of which needs to be an
invariant as well.  As the set of homogeneous invariants of fixed degree has the structure of a
vector space, one wishes to construct a basis of this vector space degree per degree in order to
obtain a generating (yet infinite) set of the invariant algebra. Although this goal has not yet
been achieved, Grassl et al. \cite{invarqubit} have constructed generating (though nonminimal) sets
of these spaces, using earlier work of Rains \cite{RainsPol}. Their study of local invariants is
general in the sense that is does not merely regard (projectors associated with) quantum codes, but
in fact arbitrary $n$-qubit density operators $\rho$. Closer inspection of their basic invariants
when dealing with stabilizer codes is certainly appropriate, given the very specific structure of
these codes and their associated projectors. Indeed, the stabilizer formalism has an equivalent
formulation in terms of algebra over GF(2) and in this framework any stabilizer code of length $n$
and dimension $k$ is essentially described by an $2n\times k$ binary matrix, called the generator
matrix of the code. It is therefore natural to ask what the structure of the basic invariants is in
relation to this binary description. In the present paper we resolve this issue: it is shown that
the invariants of Grassl et al. are in a one-to-one correspondence with the dimensions of certain
linear subspaces over GF(2) which depend solely on the binary matrix description of a code. Hence,
a complete translation of the invariants into the binary stabilizer framework is obtained.

We wish to point out that the relevance of this investigation stretches beyond the domain of
quantum coding theory: stabilizer codes with rank one projectors correspond to the class of pure
states generally known as stabilizer states.  The problem of recognizing local unitary equivalence
of stabilizer states is of importance in the study of multipartite entanglement
\cite{entgraphstate, localcliffgraph} and in the development of the one-way quantum computer, a
measurement-based model of quantum computation which uses a stabilizer state as a universal
resource \cite{1wayQC}.

%This paper is structured as follows: in section II, we fix some basic notations which will be used
%in this paper; in sections III and IV, we recall the standard formulation of the stabilizer
%formalism in terms of linear algebra over GF(2) and the construction of the set of basic invariants
%of Grassl et al., respectively. In section V the main result of this paper, i.e. the translation of
%these basic invariants in terms of the binary language, is presented and discussed. The proof of
%our main result is given in section VI. Finaly, a conclusion is formulated in section VII.

\section{Notations}

First, we fix some basic notations which will be used throughout this paper. Seeing that stabilizer
codes have descriptions both as projectors on a complex Hilbert space and as binary linear spaces,
we will be dealing with algebra over the fields $\Bbb{C}$ and $\Bbb{F}_2 :=$ GF(2), where the
latter is the finite field of two elements (0 and 1), where arithmetics are performed modulo 2. The
set of $p\times q$ matrices over a field $\Bbb{F}\in\{\Bbb{C},\Bbb{F}_2\}$ will be denoted by
$M_{p\times q}(\Bbb{F})$, where $p,q\in \Bbb{N}_0$. To shorten notations, the set of square
$p\times p$ matrices is denoted by $M_{p}(\Bbb{F})$.

The group ${\cal S}_r$ is the symmetric group of order $r$. For any $n\in\Bbb{N}_0$, ${\cal S}_r^n$
denotes the $n$-fold cartesian product of ${\cal S}_r$ with itself, i.e., ${\cal S}_r^n$ consists
of all $n$-tuples $\Pi:=(\pi_1, \dots, \pi_n)$, where $\pi_i\in {\cal S}_r$ for every $i=1, \dots,
n$.

\section{Stabilizer codes and linear spaces over GF(2)}

The Pauli group ${\cal G}_n$ on $n$ qubits consists of all $4\times 4^n$ $n$-fold tensor products
of the form $\alpha\ v_1 \otimes v_2\otimes \dots \otimes v_n$, where $\alpha \in \{ \pm 1, \pm
i\}$ is an overall phase factor and the $2 \times 2$-matrices $v_i$ $(i=1,\dots, n)$ are either the
identity $\sigma_0$ or one of the Pauli matrices \[ \sigma_x= \left(
\begin{array}{cc}0 & 1\\1 & 0 \end{array}\right),\ \sigma_y = \left(
\begin{array}{cc}0 & -i\\i & 0 \end{array}\right),\ \sigma_z=\left(
\begin{array}{cc}1 & 0\\0 & -1 \end{array}\right).  \]

An $n$-qubit stabilizer ${\cal S}$ in the Pauli group is a subgroup of ${\cal G}_n$ which is
generated by $k\leq n$ commuting, independent and Hermitian observables $M_i \in {\cal G}_n$ ($i=1,
\dots, k$). Here "independent" means that no product of the form $M_1^{x_1}\dots M_k^{x_k}$, where
$x_i \in \{0,1\}$, yields the identity except when all $x_i$ are equal to zero. The stabilizer code
associated with ${\cal S}$ is the joint eigenspace belonging to eigenvalue one of the $k$ operators
$M_i$. The numbers $n$ and $k$ are called the length and the dimension of the code, respectively.
There is a one-to-one correspondence between the code associated with ${\cal S}$ and the matrix
\begin{equation}\label{rho_S}\rho_{\cal S}=\frac{1}{2^n}\sum_{M\in {\cal S}}M,\end{equation} as this operator is (up to a
multiplicative constant) the projection operator which projects on the code space. The
normalization is chosen such as to yield Tr$(\rho_{\cal S})=1$.

We now briefly discuss the binary representation of the stabilizer formalism (for literature on
this subject, see e.g. \cite{Gott, QCQI}). Employing the mapping
\begin{eqnarray}
\sigma_0=\sigma_{00} &\mapsto& (0,0)\nonumber\\
\sigma_x=\sigma_{01} &\mapsto& (0,1)\nonumber\\
\sigma_z=\sigma_{10} &\mapsto& (1,0)\nonumber\\
\sigma_y=\sigma_{11} &\mapsto& (1,1),
\end{eqnarray}
the elements of ${\cal G}_n$ can be represented as $2n$-dimensional binary vectors as follows:
\[\sigma_{u_1v_1}\otimes\dots\otimes\sigma_{u_nv_n}=\sigma_{(u,v)} \mapsto  (u,v) \in \Bbb{F}_2^{2n},\]
where $(u,v)=(u_1, \dots, u_n, v_1, \dots,v_n)$. This parameterization establishes a group
homomorphism between ${\cal G}_n, \cdot$ and $\Bbb{F}_2^{2n},+$ (which disregards the overall
phases of Pauli operators). In this binary representation, two Pauli operators $\sigma_a$ and
$\sigma_b$, where $a, b \in \Bbb{F}_2^{2n}$, commute if and only if $a^T P b = 0$, where the
$2n\times 2n$ matrix
\[P = \left[
\begin{array}{cc} 0 & I \\ I& 0 \end{array} \right]\]
 defines a symplectic inner product on $\Bbb{F}_2^{2n}$. Therefore, a code of length $n$
and dimension $k$ corresponds to a $k$-dimensional linear subspace of $\Bbb{F}_2^{2n}$ which is
self-orthogonal with respect to this symplectic inner product, i.e., $a^T P b = 0$ for every $a, b$
in this subspace. Given a set of generators of the stabilizer, we assemble their binary
representations as the columns of a full rank $2n\times k$ matrix $S$, which is referred to as a
generator matrix of the stabilizer subspace. This generator matrix satisfies $S^TPS=0$ from the
symplectic self-orthogonality property. The entire binary stabilizer subspace (or \emph{code
space}) ${\cal C}_S$ consists of all linear combinations of the columns of $S$, i.e., it is equal
to
\begin{equation}
{\cal C}_S:=\left\{ Sx\ |\ x\in\Bbb{F}_2^{k}\right\}.
\end{equation}

It is important to notice what happens when some of the qubits in the system $\rho_{\cal S}$ are
traced out. For every $\omega\subseteq\{1,\dots,n\}$, the partial trace Tr$_{\omega^c}\ \rho_{\cal
S}=:\rho_{\cal S}^{\omega}$ yields a stabilizer code on $|\omega|$ qubits, where $\omega^c$ is the
complement of $\omega$ in $\{1,\dots,n\}$. Using the definition (\ref{rho_S}), it follows that
\begin{equation}\label{rho_omega}\rho_{\cal S}^{\omega}=\frac{1}{2^n}\sum \mbox{Tr}_{\omega^c}\ M.\end{equation}
As all three Pauli matrices have zero trace, the sum can be taken over all $M\in{\cal S}$ which are
equal to the identity $\sigma_0$ on the $i$th tensor factor for every $i\in\omega^c$. Defining the
\emph{support} supp$(M)$ of any $M=\alpha\ v_1 \otimes \dots \otimes v_n\in{\cal G}_n$ by the
subset of those $i\in\{1,\dots,n\}$ such that $v_i\neq\sigma_0$, the sum in (\ref{rho_omega}) runs
over the subgroup of all $M\in{\cal S}$ such that supp$(M)\subseteq\omega$. For such $M$, the
partial trace $\mbox{Tr}_{\omega^c}\ M$ removes the tensor factor $\sigma_0$ on positions
$i\in\omega^c$. Transferring the definition of $supp$ to the binary representation of ${\cal G}_n$,
the binary code space ${\cal C}_S^{\omega}$ of $\rho_{\cal S}^{\omega}$ is obtained by considering
the subspace of those $y\in {\cal C}_S$ such that supp$(y)\subseteq \omega$ and removing form these
$y$'s the components $(y_i,y_{n+i})=(0,0)$ for every $i\in\omega^c$.
%\[\mbox{supp}(v):=\{i\in\{1,\dots,n\}|(v_i, v_{n+i})\neq (0,0)\},\]

{\it Stabilizer states and graph states.} If the dimension of a stabilizer code is equal to its
length, i.e., if $k=n$, then the code is called self-dual. It is easy to see that the code space of
a self-dual code is one-dimensional or, equivalently, $\rho_{\cal S}= |\psi\rangle\langle\psi|$ for
some pure state $|\psi\rangle$. The states $|\psi\rangle$ obtained in this way are known in the
literature as \emph{stabilizer states}. By definition, a stabilizer state is the unique
simultaneous eigenvector with eigenvalue 1 of a set of $n$ commuting and independent Pauli
operators. A subset of the class of stabilizer states which will be of particular interest in our
investigation is constituted by the so-called \emph{graph states} \cite{1wayQC, graphbriegel}. For
these states, the defining eigenvalue equations can be constructed on the basis of a graph: when
$G$ is a simple graph on $n$ vertices with adjacency matrix $\theta$ \footnote{A simple graph $G$
has no loops or multiple edges. Therefore, it can be described by a $n\times n$ symmetric matrix
$\theta$ where $\theta_{ij}$ is equal to 1 whenever there is an edge between vertices $i$ and $j$
and zero otherwise. As $G$ has no loops, $\theta_{ii}=0$ for every $i=1, \dots, n$}, one defines
$n$ (commuting) Pauli operators
\[K_j = \sigma^{(j)}_x \prod_{k=1}^n \left(\sigma_z^{(k)}\right)^{\theta_{kj}},\] where
$\sigma^{(i)}_x, \sigma^{(i)}_y, \sigma^{(i)}_z$ are the Pauli operators which have resp.
$\sigma_x, \sigma_y, \sigma_z$ on the $i$th position in the tensor product and the identity
elsewhere. The graph state $|G\rangle$ is the stabilizer state associated with the operators $K_j$,
$j=1, \dots, n$. The rank one projector $|G\rangle\langle G|$ is denoted by $\rho_G$. Note that the
binary code space of a graph state $|G\rangle$, for a graph $G$ with adjacency matrix $\theta$, is
generated by
\[S = \left [ \begin{array}{c} \theta\\
I \end{array}\right].\]

{\it Local Clifford operations.} The Clifford group ${\cal C}_1$ on one qubit is the normalizer of
${\cal G}_1$ in $U(2)$, i.e. it is the subgroup of $2\times 2$ unitary operators which map ${\cal
G}_1$ to itself under conjugation. The local Clifford group ${\cal C}_n^l:={\cal C}_1^{\otimes n}$
on $n$ qubits is the $n$-fold tensor product of ${\cal C}_1$ with itself. When disregarding the
overall phases of the elements in ${\cal G}_1$, it is easy to see there exists a one-to-one
correspondence between the one-qubit Clifford operations and the 6 possible invertible linear
transformations of $\Bbb{F}_2^{2}$, since each one-qubit Clifford operator performs one of the 6
possible permutations of the Pauli matrices and leaves the identity fixed.  Generalizing to
$n$-qubit local Clifford operations, it follows that each $U\in {\cal C}_n^l$ corresponds to a
matrix $Q\in M_{2n}(\Bbb{F}_2)$ of the block form
\[Q = \left [ \begin{array}{cc} A&B\\
C&D \end{array}\right],\] where the $n\times n$ matrices $A, B, C, D$ are diagonal. We denote the
diagonal entries of $A, B, C, D$, respectively, by
 $a_i$, $b_i$, $c_i$, $d_i$, respectively. The $n$ submatrices \[Q^{(i)}:=\left [
\begin{array}{cc} a_i & b_i
\\ c_i& d_i\end{array} \right ]\] correspond to the tensor factors of $U$. It follows from the above discussion
that each of the matrices $Q^{(i)}$ is invertible. We denote the group of all such $Q$ by $C^l_n$.
It follows that two stabilizer codes $\rho_{\cal S}$, $\rho_{{\cal S}'}$ with generator matrices
$S$, $S'$, respectively, are equivalent under the local Clifford group if and only if there exists
$Q\in C^l_n$ such that
\begin{equation}\label{QSR1}
{\cal C}_{QS} = {\cal C}_{S'}.
\end{equation} To see this, simply note that $U\rho_{\cal S}U^{\dagger}=\rho_{{\cal S}'}$
for some $U\in {\cal C}_n^l$ if and only $U{\cal S}U^{\dagger}={\cal S}'$.

Finally, for our investigation it is important to note that any stabilizer state is locally
equivalent to a (generally nonunique) graph state under the local Clifford group
\cite{stabgraphcode}.

\section{Invariants, permutations and binary trees}

In this section, we recall the constructions of basic polynomial invariants reported in refs.
\cite{RainsPol, invarqubit}.

Let $\rho \in M_{2^n}(\Bbb{C})$ be an $n$-qubit density operator. Any homogeneous polynomial
$F(\rho)$ of degree $r$ in the entries of $\rho$ can be written as a trace
\[F(\rho)=\mbox{Tr }(A_F\cdot \rho^{\otimes r})\] for some matrix $A_F\in M_{2^{nr}}(\Bbb{C})$. To see this, simply note
that the tensor product $\rho^{\otimes r}$ contains all monomials of degree $r$ in the entries
$\rho_{ij}$.  The coefficients of these monomials in the polynomial $F$ are encoded in the entries
of $A_F$. Consequently, $F(\rho)$ is an invariant of $U(2)^{\otimes n}$ if and only if there exists
an $A_F$ such that
\begin{equation}\label{F}
[A_F, U^{\otimes r}]=0
\end{equation} for every $U\in U(2)^{\otimes n}$. Therefore, the study of invariant homogeneous
polynomials of fixed degree $r$ is transformed to the study of the algebra of matrices $A_F$ which
satisfy (\ref{F}). It was shown by Rains \cite{RainsPol} that a set of matrices which linearly
generate this algebra can be obtained in a one-to-one correspondence with the group ${\cal S}_r^n$
as follows: let $\Pi = (\mu, \nu, \xi, \dots)\in {\cal S}_r^n$ be an $n$-tuple of permutations. The
matrix $T_{\Pi} \in M_{2^{nr}}(\Bbb{C})$ is defined as the permutation matrix which acts on $({\Bbb
C}^{2^n})^{\otimes r}$ by permuting the $r$ copies of the $i$th qubit according to the $i$th
permutation of $\Pi$, i.e., $T_{\Pi}$ maps a tensor
\[ \psi_{i_1, j_1, k_1\dots ;\ i_2, j_2, k_2\dots ;\ \dots;\  i_r, j_r, k_r\dots}\in ({\Bbb
C}^{2^n})^{\otimes r}\] to \[  \psi_{i_{\mu(1)}, j_{\nu(1)}, k_{\xi(1)}\dots ;\ i_{\mu(2)},
j_{\nu(2)}, k_{\xi(2)}\dots ;\ \dots;\ i_{\mu(r)}, j_{\nu(r)}, k_{\xi(r)}\dots}.\] If $\Pi$ ranges
over all elements in ${\cal S}_r^n$, the matrices $T_{\Pi}$ linearly generate the algebra defined
by (\ref{F}). Therefore, one obtains a generating set of basic invariants $I_{r, \Pi}$ of degree
$r$, where
\begin{equation}\label{TPi} I_{r, \Pi}(\rho):= \mbox{Tr }(T_{\Pi}\cdot \rho^{\otimes r}).
 \end{equation} However, linear dependencies within the set of matrices $T_{\Pi}$ do
exist. In ref. \cite{invarqubit} Grassl et al. improved the above result, as the authors presented
a method which is able to pinpoint within the set $\{I_{r, \Pi}\}_{\Pi}$ a linearly independent
subset for every $r$. Their approach was to consider \emph{binary trees} and to associate with
every binary tree $B$ on $r$ nodes a permutation $\pi(B)\in {\cal S}_r$. Enumeration of all
possible $n$-tuples of permutations obtained in this way then yields a basis of the space of
matrices $A_F$ which satisfy (\ref{F}). We now repeat the details of this construction.

A (labelled, ordered and connected) binary tree $B$ on $r$ vertices is a special instance of a
simple, oriented and connected graph, i.e. it consists of a set of  vertices or nodes $V=\{1,
\dots, r\}$ which can be connected by arrows according to a number of prescriptions. If there is an
arrow from a node $f\in V$ to a node $s\in V$ then $f$ is called the father of $s$ and, conversely,
$s$ is a son of $f$. In a binary tree, all nodes but one have exactly one father. The one node
without father is called the root of the tree. Furthermore, every node has at most two sons (called
left and right son, respectively). The labelling of the $r$ nodes is obtained by traversing the
tree in the order root - left subtree - right subtree.
%The above definitions can most easily understood by considering an example as in Fig. 1.
A \emph{maximal right path} $p$ in a binary tree $B$ is an ordered tuple of nodes $p= (v_0, v_1,
\dots, v_s)$ such that $v_0$ is not the right son of any node of $B$, $v_{i}$ is the right son of
$v_{i-1}$ for $i=1, \dots,  s$ and $v_s$ has no right son. An example of a labelled binary tree is
given in Fig. 1.

\begin{figure}
\includegraphics{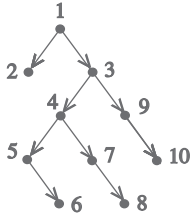}% Here is how to import EPS art
\caption{\label{fig:epsart} Binary tree on 10 nodes with maximal right paths $(1,3,9,10)$, $(2)$,
$(4,7,8)$ and $(5,6)$. Note the canonical way in which the nodes are labelled.}
\end{figure}

Denoting by ${\cal R}(B)$ the set of all maximal right paths of $B$, the permutation $\pi(B)$
associated with the binary tree $B$ is defined by the product of cycles
\[\pi(B)=\prod_{(v_0, v_1, \dots, v_s)\ \in\ {\cal R}(B)}(v_0 v_1 \dots v_s).\] Note that
$\pi(B)\in {\cal S}_r$ whenever $B$ has exactly $r$ nodes and that there is a one-to-one
correspondence between $B$ and $\pi(B)$. The set of all permutations obtained in this way is
denoted by ${\cal P}_r$. According to the result in \cite{invarqubit}, the invariants $\{I_{r,
\Pi}\}$, where $\Pi=(\pi_1, \dots, \pi_r)\in{\cal P}_r^n$ ranges over all $n$-tuples of
permutations in ${\cal P}_r$, forms a basic set of homogeneous invariants of degree $r$.

To conclude this section, we state some definitions regarding binary trees, which will be used
below. Let $B$ be a binary tree on $r$ nodes.  The \emph{start} st$(p)$ of a path $p= (v_0, v_1,
\dots, v_s)\in {\cal R}(B)$ is the element $v_0$ and the \emph{finish} fin$(p)$ is the element
$v_s$. The \emph{length} of $p$ is the number $s+1$. By an expression of the form "$i\in p$" is
meant that the node $i$ belongs to the set $\{v_0, v_1, \dots, v_s\}$ (note that, due to the
canonical labelling of the nodes, there is a one-to-one correspondence between $p$ and the set
$\{v_0, v_1, \dots, v_s\}$). For every node $i$, the path $p(i)\in{\cal R}(B)$ is the unique
maximal right path such that $i\in p(i)$.

Let $B$ have $t:=|{\cal R}(B)|$ maximal right paths $p_1, \dots, p_t$, which we suppose to be
ordered in such a way that st$(p_1)$ $<$ st$(p_2)$ $< \dots <$ st$(p_t)$. The columns $(R_B)_j$ of
the  matrix $R_B\in M_{r\times t}(\Bbb{F}_2)$ are defined by:
\begin{equation}\label{RB}
(R_B)_j = \sum_{i\in p_j} e_i,
\end{equation}
for every $j\in\{1,\dots, t\}$, where $e_i$ is the $i$th canonical base vector in $\Bbb{F}_2^r$.
The columns $(D_B)_j$ of the matrix $D_B\in M_{r}(\Bbb{F}_2)$ are defined by:
\begin{equation}
(D_B)_j = \sum_{i\in p(j),\ i\leq j} e_i,
\end{equation}
for every $j\in\{1,\dots, r\}$. Finally, the linear space $V_{B}$ consists of all $x\in\Bbb{F}_2^r$
such that $\sum_{i\in p}x_i=0$ for every $p\in {\cal R}(B)$ (i.e., $V_{B}$ is the null space of the
matrix $R_B^T$).
%The matrix $R_{B}\in M_{r\times t}(\Bbb{F}_2)$ is then defined by
%\begin{equation}\label{RB}(R_{B})_{ij} = \left\{
%\begin{array}{cl} 1 &\mbox{ if } i \in p_j\\ 0 & \mbox{ else } \end{array}\right. \end{equation}
%Note that, due to the canonical labelling of the nodes, there is a one-to-one correspondence
%between $B$ and the set ${\cal R}(B)$ of its maximal right paths. note that any $i\in\{1, \dots,
%r\}$ belongs to exactly one maximal right path of $B$.

\section{Main result and discussion}

%In this section we state the central result of this paper, which relates the above basis of
%invariants, which are defined as functions of the set of $n$-qubit density operators, to objects
%which can entirely be formulated in terms of the binary representation of
We are now in a position to state the central result of this paper.

\textbf{Theorem 1} {\it Let $\rho_{\cal S}$ be a stabilizer code of length $n$ and dimension $k$
with generator matrix $S$. Let $S_i^T$ ($i=1, \dots, n$) be the $2\times k$ submatrix of $S$
obtained by selecting the $i$th and the $(n+i)$th row of $S$. Fix $r\in\Bbb{N}_0$, let $B_1, B_2,\
\dots, B_n$ be $n$ binary trees on $r$ nodes and let $\Pi\in{\cal P}_r^n$ be the associated
$n$-tuple of permutations. Then
\begin{equation}\label{logI=dimV}
log_2\ I_{r,\Pi}(\rho_{\cal S}) \sim \mbox{dim}_{\Bbb{F}_2}\mbox{ ker } \left[ \begin{array}{c} R_{B_1}^T\otimes S_1^T \\
R_{B_2}^T\otimes S_2^T\\ \dots \\R_{B_n}^T\otimes S_n^T\end{array}\right],
\end{equation}
where $\sim$ denotes equality up to an additive constant independent of $\rho_{\cal S}$. }

%Let $G$ be a graph with adjacency matrix $\theta \in \Theta$ and let $\overrightarrow \pi
%=(\mu_1,\dots, \mu_n)$ be an $n$-tuple of permutations $\mu_i\in {\cal P}_r$. Let the vector space
%$V_{r,\overrightarrow \pi}(G)$ consist of all binary $n\times r$-matrices
%$X=(x^{(1)}|\dots|x^{(r)})$, with columns $x^{(j)}$, which are solutions to the following
%equations:
%\[S_{i} \left\{\sum_{j\in \mbox{ \scriptsize{supp}}({\cal C}_i^k) }x^{(j)} \right\}=0, \mbox{\quad
%for all } i=1, \dots, n,\ \mbox{ for all } k,
%\] where $S_{i} = \left[\begin{array}{c} \theta_i^T \\ e_i^T
%\end{array}\right]$ and the ${\cal C}_i^k$'s are the different cycles of $\mu_i$. Then
%\begin{equation}\label{logI=dimV}
%\log_2 I_{r,\overrightarrow \pi}(|G\rangle\langle G|) \sim\mbox{dim } V_{r,\overrightarrow \pi}(G),
%\end{equation}

Theorem 1 shows how the information contained in the invariants $I_{r,\Pi}$ can be recuperated
within the binary representation of the stabilizer formalism.  The fact that \emph{a} translation
into the binary framework is possible is of course not unexpected, as a stabilizer code is, up to
information about the overall phases of its stabilizer elements, defined by its generator matrix.
Moreover, these phases do not play a role in determining the (local) equivalence class of a code
\cite{localcliffgraph}. However, the simple form of the result (\ref{logI=dimV}) is remarkable: an
invariant $I_{r,\Pi}$ is in a one-to-one correspondence with the dimension of a binary linear space
which depends only on the generator matrix $S$ - and this in a very transparant way. Additionally,
it is interesting to notice the explicit way in which the $n$-tuple of binary trees appear in the
result: every matrix $R_{B_i}$, corresponding to the $i$th binary tree, is coupled via a tensor
product to the matrix $S_i$, which is the subblock of $S$ containing the information about the
$i$th qubit. Finally, we note that the r.h.s. of (\ref{logI=dimV}) can be computed efficiently via
a calculation of the rank over $\Bbb{F}_2$ of the matrix
\begin{equation}\label{matrix}
\left[ \begin{array}{c} R_{B_1}^T\otimes S_1^T \\
R_{B_2}^T\otimes S_2^T\\ \dots \\R_{B_n}^T\otimes S_n^T\end{array}\right].\end{equation}

%Indeed, it follows from (\ref{logI=dimV}) that the $I_{r,\Pi}$'s, which are defined as functions on
%the set of states $\rho_{\cal S}$, can be reformulated as functions on the set of generator
%matrices.

Before proving theorem 1 in section VI, we investigate the invariants (\ref{logI=dimV}) in more
detail. We start with the invariants of smallest nontrivial degree, i.e. $r=2$. There are exactly
two binary trees on 2 vertices, as node 2 can either be the right or the left son of node 1.
Equivalently, there are two possible matrices $R_{B}$ according to definition (\ref{RB}), namely
\begin{equation}\label{degree2}
\left[\begin{array}{cc}1&0\\0&1\end{array} \right] \mbox{\quad and \quad } [1\ 1]^T,
\end{equation}
where the identity matrix corresponds to the tree where 2 is the left son of 1. Now, consider an
$n$-tuple $(B_1, \dots, B_n)$ of binary trees and the corresponding $n$-tuple of permutations
$\Pi\in {\cal P}_2^n$. Let $\omega\subseteq\{1,\dots,n\}$ denote the set of all $i$ such that
$R_{B_i}= [1\ 1]^T$ - note that every $n$-tuple of binary trees on 2 nodes corresponds uniquely to
such a set $\omega$. Using the notation $S_i$ as in theorem 1, this implies that \[R_{B_i}^T\otimes
S_i^T = [S_i^T \ S_i^T ]\]
 whenever
$i\in \omega$ and \[R_{B_i}^T\otimes S_i^T = \left[\begin{array}{cc}S_i^T&0\\0&S_i^T\end{array}
\right]\] otherwise. Therefore, the null space of the matrix (\ref{matrix}) consists of all vectors
$(x, x')\in
\Bbb{F}_2^{2k}$ such that \begin{eqnarray}\label{degree2'}&S_i (x+x')=0& \mbox{ for every } i\in \omega\nonumber\\
&S_jx=S_jx'=0& \mbox{ for every }j\in\omega^c , \end{eqnarray} where $\omega^c$ is the complement
of $\omega$ in $\{1, \dots, n\}$. Note that (\ref{degree2'}) implies that $Sx=Sx'$ and therefore
$x=x'$, since $S$ has full rank. Thus, the solutions of (\ref{degree2'}) are in a one-to-one
correspondence with the linear subspace of $\Bbb{F}_2^{k}$ of those $x$ satisfying $S_jx=0$ for
every $j\in\omega^c$. The linear mapping $\phi_S: \Bbb{F}_2^{k}\to \Bbb{F}_2^{2n}$ defined by the
matrix $S$ maps the space of such $x$'s to the space of vectors $y=Sx$ which satisfy
$y_{j}=y_{n+j}=0$ for every $j\in \omega^c$. As $S$ has full rank, the mapping $\phi_S$ is
injective and the spaces of the $x$'s and the $y$'s have equal dimension. Recalling that the
\emph{support} supp$(v)$ of any $v\in\Bbb{F}_2^{2n}$ is the subset of those $i\in\{1,\dots,n\}$
such that $(v_i, v_{n+i})\neq (0,0)$,
%\[\mbox{supp}(v):=\{i\in\{1,\dots,n\}|(v_i, v_{n+i})\neq (0,0)\},\]
we can state that the supports of the $y$'s lie within the set $\omega$. Thus, we have shown that
\begin{equation}\label{suppkleiner} \log_2 I_{2, \Pi} \sim \mbox{ dim }\{y\in {\cal C}_S| \mbox{ supp}(y)\subseteq
\omega \}.\end{equation} This is clearly a more insightful presentation of the invariants $I_{2,
\Pi}$ than (\ref{logI=dimV}), as (\ref{suppkleiner}) relates invariants to the dimensions of the
subspaces ${\cal C}_S^{\omega}$ of the code space ${\cal C}_S$. Moreover, a similar argument as
above can be made to obtain an analogous presentation of the invariants of higher degree:

\textbf{Theorem 2} {\it Let $S$ be a $2n\times k$ generator matrix of a stabilizer code and let the
subblocks $S_i^T$ be defined as in theorem 1. Fix $r\in\Bbb{N}_0$ and let $B_1, \dots, B_n$ be $n$
binary trees on $r$ vertices. Let ${\cal R}= {\cal R}(B_1)\cup\dots\cup{\cal R}(B_n)$ denote the
set of all maximal right paths of these trees. For every $p\in{\cal R}$, let $\omega_{p}$ denote
the subset of all $i\in\{1,\dots,n\}$ such that $p\notin{\cal R}(B_i)$. Then the dimension of the
kernel of the matrix (\ref{matrix}) is equal to the dimension of the space
%\begin{equation}\label{theorem2}
%\{ (y^{(1)},\dots, y^{(r)})\in\ {\cal C}_S\times\dots\times\ {\cal C}_S |\
%  supp\ (\sum_{j\in p}y^{(j)})\subseteq\omega_{p},\  for\ every\ p\in{\cal
% R}\}.
%\end{equation}}
\begin{eqnarray}\label{theorem2}
&&\{ (y^{(1)},\dots, y^{(r)})\in\ {\cal C}_S\times\dots\times\ {\cal C}_S |\quad
 \nonumber\\&&\mbox{\quad \quad \quad} supp\ (\sum_{j\in p}y^{(j)})\subseteq\omega_{p},\  for\ every\ p\in{\cal
 R}\}.\nonumber\\
\end{eqnarray}}
The proof of theorem 2 is omitted, as it is a straightforward generalization of the considerations
made above for the invariants of degree 2. A number of properties of the invariants immediately
follow from theorem 2: e.g., consider an invariant $I_{r+1, \Pi_0}$ such that its $n$ binary trees
$B_i$ have got a common maximal right path $p_0$ of exactly one element $i_0$, i.e.
$p_0=(i_0)\in{\cal R}(B_1)\cap\dots\cap{\cal R}(B_n)$. Considering (\ref{theorem2}) for this
invariant, it follows that $\omega_{p_0}=\emptyset$. Consequently, supp$(y^{(i_0)}) \subseteq
\emptyset$ and therefore $y^{(i_0)}=0$ for every $y^{(i_0)}$ in (\ref{theorem2}). Thus, the
invariant $I_{r+1, \Pi_0}$ is equal to an invariant of degree $r$ corresponding to binary trees
$\bar B_i$ which are obtained by deleting the node $i_0$ from the original trees $B_i$. Reversing
the above argument shows that any invariant of degree $r$ can be written as an invariant of degree
$r+1$, thereby showing that
\[\{I_{r, \Pi}\}_{\Pi}\subset\{I_{r+1, \Pi'}\}_{\Pi'}\] for every $r$, where $\Pi$  ($\Pi'$) ranges
over all elements in ${\cal P}_r^n$ (${\cal P}_{r+1}^n$). Moreover, this argument can be
generalized to show that an invariant corresponding to a tuple of binary trees which have some
subtree in common, can be written as the product of invariants of lower degree.

%\subsection*{Discussion}

\section{Proof of Theorem 1}
The proof of theorem 1 is given in two main parts. We show in subsection A that it is sufficient to
prove the theorem for graph states. The proof of theorem 1 for graph states is subsequently given
in subsection B. For the remainder of this section, we fix $r$ and consider an $n$-tuple
$\Pi=(\pi_1, \dots, \pi_n)$ of permutations $\pi_i\in {\cal P}_r$ and $n$ binary trees $B_i$ on $r$
vertices corresponding to the permutations $\pi_i$. We denote ${\cal B}:=(B_1, \dots, B_n)$.

\subsection{Reduction to graph states}

Suppose that theorem 1 holds for all graph states. Let $\rho_{\cal S}$ be a stabilizer code of
length $n$ and dimension $k$ with generator matrix $S$. For large enough $m>n$, there exists a
stabilizer state $|\psi\rangle$ on $m$ qubits such that $\rho_{\cal S}$ can be obtained from
$|\psi\rangle$ by tracing out the qubits $n+1,\dots, m$, i.e.
\[\rho_{\cal S}=|\psi\rangle\langle\psi|^{\{1,\dots,n\}}\] using the notation of section 3.
Furthermore, $|\psi\rangle$ is equivalent to some graph state $|G\rangle$ under the local Clifford
group. Denoting $\omega=\{1,\dots,n\}$, it follows that $\rho_{\cal S}$ is locally equivalent to
$\rho_G^{\omega}$. Letting $S'$ be the generator matrix of $|G\rangle$, this last fact translates
into the binary picture as \begin{equation}\label{reduction0}{\cal C}_{QS} = {\cal
C}_{S'}^{\omega}\end{equation} for some $Q\in C^l_n$. Now, let $\Pi'\in{\cal P}_r^m$ be the
$m$-tuple of permutations which is obtained by appending to $\Pi$ $m-n$ times the identity
permutation (which belongs to ${\cal P}_r$) and let ${\cal B}'=(B_1,\dots, B_n, B_0,\dots, B_0)$ be
the associated $m$-tuple of binary trees; here $B_0$ is the binary tree with $r$ maximal right
paths $(i)$, corresponding to the identity permutation. The crucial observation is now
\[I_{r,\Pi}(\rho_{\cal S})= I_{r,\Pi'}(\psi).\] This identity can easily be verified by using the
definition of $I_{r,\Pi'}$. Moreover, $ I_{r,\Pi'}(\psi)= I_{r,\Pi'}(\rho_G)$ since the states
$|\psi\rangle$ and $|G\rangle$ are locally equivalent.  We can now apply theorem 1 and find

\begin{eqnarray}\label{reduction1}
\log_2 I_{r,\Pi}(\rho_{\cal S})&=& \log_2 I_{r,\Pi'}(\rho_G)\nonumber\\
&\sim& \mbox{dim}_{\Bbb{F}_2}\mbox{ ker } \left[ \begin{array}{c} R_{B_1}^T\otimes S_1^{'T}
\\\dots\\R_{B_n}^T\otimes S_n^{'T}\\ R_{B_0}^T\otimes S_{n+1}^{'T}\\ \dots \\R_{B_0}^T\otimes S_m^{'T}\end{array}\right]
\end{eqnarray}
Applying theorem 2 to the generator matrix $S'$ and the binary trees ${\cal B}'$,
(\ref{reduction1}) is equal to the dimension of
\begin{eqnarray}\label{reduction2}
&&\{ (y^{(1)},\dots, y^{(r)})\in\ {\cal C}_{S'}\times\dots\times\ {\cal C}_{S'} |\quad
 \nonumber\\&&\mbox{\quad \quad \quad} \mbox{supp}\ (\sum_{j\in p}y^{(j)})\subseteq\omega_{p},
 \mbox{ for every } p\in {\cal
 R}'\}\nonumber\\
\end{eqnarray}
where ${\cal  R}'={\cal R}(B_1)\cup\dots\cup{\cal R}(B_n)\cup{\cal R}(B_0)$. As the last $m-n$
trees in the $m$-tuple ${\cal B}'$ are equal to $B_0$, every $y^{(j)}$ in (\ref{reduction2}) has
supp$(y^{(j)})\subseteq\omega$. Therefore, the dimension of (\ref{reduction2}) is equal to the
dimension of
\begin{eqnarray}\label{reduction3}
&&\{(x^{(1)},\dots, x^{(r)})\in\ {\cal C}_{S'}^{\omega}\times\dots\times\ {\cal C}_{S'}^{\omega}
|\quad
 \nonumber\\&&\mbox{\quad \quad \quad} \mbox{supp}\ (\sum_{j\in p}x^{(j)})\subseteq\omega_{p},\mbox{ for every }  p\in{\cal
 R}\}\nonumber\\
\end{eqnarray}
where now ${\cal R}={\cal R}(B_1)\cup\dots\cup{\cal R}(B_n)$. Finally, we recall the identity
(\ref{reduction0}) and note that (\ref{reduction3}) remains invariant if ${\cal C}_{QS}$ is
replaced by ${\cal C}_{S}$. A last application of theorem 2 yields
\begin{equation}
\log_2 I_{r,\Pi}(\rho_{\cal S}) \sim \mbox{dim}_{\Bbb{F}_2}\mbox{ ker } \left[ \begin{array}{c} R_{B_1}^T\otimes S_1^T \\
R_{B_2}^T\otimes S_2^T\\ \dots \\R_{B_n}^T\otimes S_n^T\end{array}\right],
\end{equation}
which is the desired result.

%\[I_{r, \Pi}(\Psi) = \sum \ \Psi_{i_{\mu(1)}j_{\nu(1)}k_{\xi(1)}\dots
%}^{i_1j_1k_1\dots}\ \Psi_{i_{\mu(2)}j_{\nu(2)}k_{\xi(2)}\dots }^{i_2j_2k_2\dots} \dots\]

\subsection{Proof of theorem 1 for graph states}

Fix a  graph $G$ on $n$ vertices with adjacency matrix
$\theta$ and the generator matrix \[S = \left [ \begin{array}{c} \theta\\
I \end{array}\right],\] which has $2\times n$ subblocks $S_i^T$ defined as in theorem 1 by
\[S_i^T = \left[
\begin{array}{c} \theta_i^T\\e_i^T
\end{array}\right].\]
The proof of theorem 1 for the case where $\rho_{\cal S}=\rho_G$ is structured as follows: in lemma
3 we show that the invariant $I_{r,\Pi}(\rho_{G})$ is equal to a sum of the form
\begin{equation}\label{sketch} \frac{1}{N}\sum_{X \in {\cal V}} (-1)^{{\cal
Q}(X)},\end{equation} where $N$ is a normalization factor independent of $G$, ${\cal V}$ is a
linear subspace of $M_{n\times r}(\Bbb{F}_2)$ and ${\cal Q}$ is a quadratic form on $M_{n\times
r}(\Bbb{F}_2)$. Preliminary material used to prove this result will be gathered in lemmas 1 and 2.
In lemma 4, we subsequently show that the form ${\cal Q}$ is in fact identical zero on the space
${\cal V}$, which implies that the sum in (\ref{sketch}) is (up to the normalization) equal to the
cardinality of ${\cal V}$. Finally, this cardinality is related to the r.h.s of (\ref{logI=dimV})
and the proof of theorem 1 is completed.

It will be convenient to work with a real variant of the set of Pauli matrices (as in ref.
\cite{stab_clif_GF2}), defined by
\begin{eqnarray}
\tau_{00}&=&\sigma_{00},\nonumber\\
\tau_{01}&=&\sigma_{01},\nonumber\\
\tau_{10}&=& \sigma_{10},\nonumber\\
\tau_{11}&=& i\sigma_{11} = \left(
\begin{array}{cc}0 & 1\\-1 & 0 \end{array}\right),
\end{eqnarray}
which we will call the tau matrices. Analogous to the notation introduced in section 3, $n$-fold
products of tau matrices are represented as
\[\tau_{u_1v_1}\otimes\dots\otimes\tau_{u_nv_n}=\tau_{(u,v)},\] where $(u,v)=(u_1, \dots, u_n, v_1,
\dots,v_n)\in \Bbb{F}_2^{2n}$. We now prove a useful parameterization of the projector $\rho_G$:

%In theorem 1, the link between the Hilbert space and the binary description of graph states becomes
%very apparent.  It is shown that a graph state can be written as a sum of matrices
%$\pm\tau_{(x,y)}$ where the vector $(x,y)\in \Bbb{Z}_2^{2n}$ ranges over the binary stabilizer
%subspace of the state; furthermore, each reduced density operator can be written as a similarly
%sum, where the index vector now ranges over a subspace of the stabilizer space. This result will
%prove very useful in section III, when we will calculate LU-invariants and relate these invariants
%to graph theoretical objects.

\textbf{Lemma 1} {\it The  projector $\rho_G$ can be parameterized as follows:
\begin{equation}\label{Gsom}
\rho_{G} = \frac{1}{2^n}\sum_{x \in \Bbb{F}_2^n} (-1)^{k_{\theta}(x)}\ \tau_{(\theta x,x)},
\end{equation}
where $k_{\theta}$ is the quadratic form over $\Bbb{F}_2$ associated with $\theta$, i.e.,
$k_{\theta}(x) = \sum_{i<j}\theta_{ij}x_ix_j$. }
%(ii) For a set $\omega\subseteq\Omega$, let $\rho_{\omega}(G)$ be the reduced density matrix which
%describes the subsystem of $|G\rangle$ consisting of the qubits $i \in \omega$, i.e.
%$\rho_{\omega}(G) = \mbox{Tr}_{\bar\omega}|G\rangle\langle G|$. Define $S_{\omega}$ to be the
%$2|\omega|\times n$-submatrix of $S=\left[
%\begin{array}{c} \theta\\I\end{array}\right]$ which is obtained by retaining the rows on positions $i$ and $i+n$
%for all $i\in\omega$. Then
%\begin{equation}\label{rhosom}
%\rho_{\omega}(G) = \frac{1}{2^{|\omega|}}\sum(-1)^{x^T lows(\theta)x}\ \tau_{S_{\omega}x},
%\end{equation}
%where the sum is taken over all $x\in \Bbb{Z}_2^n$ that lie in the subspace $S_{\bar\omega}x=0$.
%Moreover, $\rho_{\omega}$ is proportional to a projection operator. Finally, let $\theta_{\omega}$
%be the off-diagonal $|\omega|\times (n-|\omega|)$-submatrix of $\theta$ which is obtained by
%retaining the rows on positions $i \in \omega$ and, within these rows, removing all the columns on
%positions $i \in \omega$. Then $\log_2 (\mbox{rank }\rho_{\omega}) =
% \mbox{rank } \theta_{\omega}$.  }

{\it Proof:} the state $|G\rangle$ is defined by the $n$ relations $\tau_{(\theta_j, e_j)}|G\rangle
= |G\rangle$, where $\theta_j$ is the $j$th column of $\theta$ and $e_j$ is the $j$th canonical
basis vector of $\Bbb{F}_2^n$. The stabilizer of $|G\rangle$ consists of all products \[M_x =
\prod_{j=1}^n {\tau_{(\theta_j, e_j)}}^{x_j},\] where $x = (x_1, \dots, x_n) \in \Bbb{F}_2^n$.
After a repeated application of the multiplication rule \cite{stab_clif_GF2}
\[\tau_{(u,v)}\tau_{(u', v')} = (-1)^{v^Tu'}\tau_{(u+u',v+v')},\] where $u, u', v, v' \in \Bbb{F}_2^n$, we arrive at
\[M_x = (-1)^{k_{\theta}(x)}\ \tau_{(\theta x,x)}.\] Since $\rho_G =
\frac{1}{2^n}\sum_{x \in \Bbb{F}_2^n} M_x$, we obtain the result. \hfill $\square$

Lemma 1 will be used below to compute the invariant $I_{r, \Pi}(\rho_G)$. After plugging
(\ref{Gsom}) in (\ref{TPi}), we will be dealing with expressions of the form
\begin{equation}\label{expr}\mbox{Tr }(T_{\Pi}\ \tau_1\otimes\tau_2\otimes\dots\otimes\tau_{r}),\end{equation}
where the $\tau_i$'s are themselves $n$-fold tensor products of the tau matrices, i.e. $\tau_i\in
{\cal G}_n$ for every $i=1, \dots, r$. A closer look at these expressions beforehand is
appropriate. To this end, let $X, Y, \dots$ be any $r$ operators in $M_{2}(\Bbb{C})^{\otimes n}$,
i.e.,
\begin{eqnarray}X&=&X_1 \otimes\dots\otimes X_n,\nonumber\\
Y&=&Y_1 \otimes\dots\otimes Y_n,\ \dots,\nonumber\end{eqnarray} where $X_i, Y_i, \dots\in
M_2(\Bbb{C})$. Furthermore, for any $\pi\in{\cal P}_r$ we denote
\[A_{r,\pi}(U, V, \dots) := \sum_{i_1, \dots, i_r }U_{i_1i_{\pi(1)}}\ V_{i_2i_{\pi(2)}}
\dots,\] where  $U$, $V, \dots$ are $r$ arbitrary  $2\times 2$ matrices. Using the definition of
$T_{\Pi}$, it is then easy to check that
\begin{equation}\label{A_pi}\mbox{Tr }(T_{\Pi}\ X\otimes Y\otimes \dots) = \prod_{i=1}^n A_{r,\pi_i}(X_i, Y_i,
\dots).\end{equation} It follows that (\ref{expr}) is a product of $n$ factors of the form
\[A_{r,\pi}(\tau_{u_1v_2}, \dots, \tau_{u_rv_r})=:A_{r,\pi}(u,v),\] where $u=(u_1,\dots,u_r)$
and $v=(v_1,\dots,v_r) \in \Bbb{F}_2^r$. Expressions of this type are calculated in lemma 2:

\textbf{Lemma 2} {\it Let $\pi \in {\cal P}_r$ be a permutation corresponding to a binary tree $B$
and let $(u,v)\in\Bbb{F}_2^{2r}$. Let the matrix $D_B$ and the space $V_B$ be defined as in section
4. Then
\[A_{r,\pi}(u,v)=
\left\{ \begin{array}{cc}2^{r -dim\ V_{B}}(-1)^{u^TD_{B}^Tv} & \mbox{ if } u,v \in V_{B}\\
0 &\mbox{ otherwise}\end{array} \right.\]}

 {\it Proof:} First, note that the entries of the
1-qubit operators $\tau_{ab}$ can be parameterized as
\[\left(\tau_{ab}\right)_{x,y} = (-1)^{a(b+x)}\delta_{x+y,b},\] where $a, b, x,y\in\Bbb{F}_2$. Using this
formula in the definition of $A_{r,\pi}$, we obtain \[A_{r,\pi}(u,v) = \sum_{x\in \Bbb{F}_2^r}
(-1)^{u^T(v+x)}\delta_{x_1+x_{\pi(1)},v_1}\dots\delta_{x_r+x_{\pi(r)},v_r}.\]

Equivalently, the sum runs over all $x\in\Bbb{F}_2^r$ which lie in the affine subspace determined
by the equations $x_i + x_{\pi(i)}=v_i$ for all $i=1, \dots, r$. However, this system of equations
may not be consistent: indeed, one can easily show that a solution exists iff $v\in V_{B}$.
Whenever this is the case, the solutions are given by $x=x_0 + x'$, where $x_0=D_{B}^Tv+v$ and $x'$
satisfies $x'_i + x'_{\pi(i)}=0$ for every $i=1, \dots, r$. Moreover, the space of all such $x'$ is
the orthogonal complement of $V_{B}$ with respect to the standard inner product in $\Bbb{F}_2^r$
(or equivalently, the column space of $R_B$), as one can verify. Therefore, we have
\[A_{r,\pi}(u,v)=
\left\{ \begin{array}{cc}\sum_{x'\in V_{B}^{\perp}}(-1)^{u^T(D_{B}^Tv + x')} & \mbox{ if } v \in V_{B}\\
0 &\mbox{ otherwise}\end{array} \right.\] Furthermore, the sum $\sum_{x'\in V_{B}^{\perp}}(-1)^{u^T
x'}$ is equal to $2^{r - \scriptsize{\mbox{dim}}\ V_{B}}$ if $u\in V_{B}$ and zero otherwise. This
proves the result. \hfill $\square$

We now proceed in calculating the invariant $I_{r,\Pi}(\rho_{G})$. Using lemma 1 in (\ref{TPi}), we
find that $I_{r,\Pi}(\rho_{G})$ is equal to the sum
%\begin{equation}\label{A1}
%  \frac{1}{2^{nr}}\sum_{x^{(1)},\ \dots,\ x^{(r)}\ \in\ \Bbb{F}_2^n}
%\left\{ (-1)^{\sum_{i=1}^r k_{\theta}(x^{(i)})} \mbox{ Tr }(T_{\Pi}\ \tau_{(\theta
%x^{(1)},x^{(1)})}\otimes\dots\otimes \tau_{(\theta x^{(r)},x^{(r)})})\right\}
%\end{equation}
\begin{eqnarray}\label{A1}
  &&\frac{1}{2^{nr}}\sum_{x^{(1)},\ \dots,\ x^{(r)}\ \in\ \Bbb{F}_2^n}
\left\{ (-1)^{\sum_{i=1}^r k_{\theta}(x^{(i)})}\times\right.\nonumber\\
&&\qquad\qquad\left.\mbox{ Tr }(T_{\Pi}\ \tau_{(\theta
x^{(1)},x^{(1)})}\otimes\dots\otimes \tau_{(\theta x^{(r)},x^{(r)})})\right\}\nonumber\\
\end{eqnarray}
First, denoting by ${\cal P}_{ \mbox{\scriptsize lows}}(\theta)$ the strictly lower triangular part
of $\theta$ and writing
\[X:=[x^{(1)}|\dots|x^{(r)}]\in M_{n\times r}(\Bbb{F}_2),\] we obtain the shorthand notation
\[\sum_{i=1}^r k_{\theta}(x^{(i)}) = \mbox{Tr }X^T {\cal P}_{ \mbox{\scriptsize lows}}(\theta)X\]

Secondly, the trace in (\ref{A1}) splits into a product of $n$ factors as in (\ref{A_pi}), each of
which can be calculated by employing lemma 2. The calculation is straightforward. Defining for
every $X\in M_{n\times r}(\Bbb{F}_2)$ the matrix $X_{\cal B}\in M_{n\times r}(\Bbb{F}_2)$ by
\[(X_{\cal B})_{ij}= \sum_{k=1}^r X_{ik}\left(D_{B_i}\right)_{kj},\] one finds that (\ref{A1}) is equal
(up to a normalization independent of $G$) to
\begin{equation}
\sum (-1)^{\scriptsize{\mbox{ Tr }} X^T {\cal P}_{ \mbox{\scriptsize lows}}(\theta)X +
\scriptsize{\mbox{ Tr }} X_{\cal B}^T\theta X},
\end{equation}
where the sum runs over all $X$ such that
\begin{equation}\label{A6}
S_i^T \left(\sum_{j\in p}x^{(j)}\right)=0
\end{equation}
for every $i\in\{1,\dots, n\}$ and $p\in{\cal R}(B_i)$. We will denote the space of all such $X$ by
$V_{\cal B}(G)$. We have proven:

%the vector $(x^{(1)},\dots,x^{(r)})\in \Bbb{F}_2^{nr}$ lies in the kernel of the matrix
%\begin{equation}
%\left[ \begin{array}{c} R_{B_1}^T\otimes S_1^T \\
%R_{B_2}^T\otimes S_2^T\\ \dots \\R_{B_n}^T\otimes S_n^T\end{array}\right],\end{equation} which is
%defined as in theorem 1 for \[S = \left [ \begin{array}{c} \theta\\
%I \end{array}\right].\]

\textbf{Lemma 3} {\it The invariant $I_{r,\Pi}(\rho_{G})$ can be written as
\begin{equation}\label{A3}
I_{r,\Pi}(\rho_{G}) =  \frac{1}{N}\sum_{X \in V_{{\cal B}}(G)} (-1)^{\scriptsize{\mbox{ Tr }} X^T
{\cal P}_{ \mbox{\scriptsize lows}}(\theta)X + \scriptsize{\mbox{ Tr }} X_{\cal B}^T\theta X},
\end{equation}
where $N$ is a normalization factor independent of $G$ and the definitions of $V_{{\cal B}}(G)$ and
$X_{\cal B}$ are as above.}

The last part of our argument consists of showing that the quadratic form ${\cal
Q}(X):={\small\mbox{ Tr }} X^T {\cal P}_{ \mbox{\scriptsize lows}}(\theta)X  + \mbox{ Tr } X_{{\cal
B}}^T\ \theta X$ is zero on the space $V_{\cal B}(G)$. Once this result is shown, the proof of
theorem 1 is immediate: indeed, if ${\cal Q}(X)=0$ for every $X\in V_{\cal B}(G)$ then $\log_2
I_{r,\Pi}(\rho_{G}) \sim \mbox{dim } V_{\cal B}(G)$. Moreover, the matrices
$X=[x^{(1)}|\dots|x^{(r)}]\in V_{\cal B}(G)$ can be reshaped as vectors $\tilde X =
(x^{(1)},\dots,x^{(r)})\in\Bbb{F}_2^{nr}$ which are exactly the elements in the null space of the
matrix
\begin{equation}
\left[ \begin{array}{c} R_{B_1}^T\otimes S_1^T \\
R_{B_2}^T\otimes S_2^T\\ \dots \\R_{B_n}^T\otimes S_n^T\end{array}\right].\end{equation} Clearly,
the spaces of the $X$'s and the $\tilde X$'s have the same dimension and the proof of theorem 1 is
thus completed. We now show that ${\cal Q}=0$ on the space $V_{\cal B}(G)$:

 \textbf{Lemma 4} {\it ${\cal Q}(X)=0$ for every $X\in V_{\cal B}(G)$.}

{\it Proof:} Let $X=[x^{(1)}|\dots|x^{(r)}]$ be an element of $V_{{\cal B}}(G)$. Recall that by
definition (\ref{A6}) this entails that
\begin{equation}\label{recall_VB}
\left[
\begin{array}{c} \theta_i^T\\e_i^T \end{array}\right]\left(\sum_{j\in p}x^{(j)}\right)=0
\end{equation}
for every $i\in\{1,\dots, n\}$ and $p\in{\cal R}(B_i)$. In particular,
\begin{equation}\label{A7}\sum_{j\in p}x^{(j)}_i =0\end{equation} for every $i\in\{1,\dots, n\}$
and for every $p\in{\cal R}(B_i)$, where $x^{(j)}=(x^{(j)}_1, \dots, x^{(j)}_n)$. Consequently,
\begin{equation}\label{sum_x_j}\sum_{j=1}^r x^{(j)}=0.\end{equation}
Now, consider the first term of ${\cal Q}(X)$:
\begin{eqnarray}\label{A4}
{\cal Q}_1&:=& \mbox{Tr }X^T {\cal P}_{ \mbox{\scriptsize lows}}(\theta)X\nonumber\\ &=&
\sum_{j=1}^r {x^{(j)}}^T {\cal P}_{ \mbox{\scriptsize lows}}(\theta)\ x^{(j)}.
\end{eqnarray}
Substituting $x^{(r)}=\sum_{j=1}^{r-1} x^{(j)}$ (from (\ref{sum_x_j})), an easy calculation shows
that
\[{\cal Q}_1 = \sum_{j=1}^{r-1}\left(\sum_{k=1}^{j-1}x^{(k)}\right)^T\theta\ x^{(j)}.\] Let
$\omega_{ij}\subseteq\{1,\dots,n\}$ consist of all $k\in\{1, \dots, j-1\}$ which belong to a
maximal right path $p$ of $B_i$ such that fin$(p)\geq j$. Then, denoting $y^{(j)}:=
\sum_{k=1}^{j-1}x^{(k)}$, (\ref{A7}) implies that $y^{(j)}_i = \sum_{k\in\omega_{ij}}x^{(k)}_i$.

The second term of ${\cal Q}(X)$ is
\begin{eqnarray}\label{A5}
{\cal Q}_2 &:=& \mbox{ Tr } X_{\cal B}^T\ \theta X\nonumber\\ &=& \sum_{j=1}^r {z^{(j)}}^T\theta\
x^{(j)},
\end{eqnarray}

where $z^{(j)}$ is the $j$th column of $X_{\cal B}$. Let $\eta_{ij}$ consist of all
$k\in\{1,\dots,j-1\}$ which belong to the unique maximal right path of $B_i$ which contains $j$. It
then follows from the definition of $X_{\cal B}$ that $z^{(j)}_i =
\sum_{k\in\eta_{ij}\cup\{j\}}x^{(k)}_i$. Note that $\eta_{ir}\cup \{r\}\in{\cal R}(B_i)$ and
therefore $z^{(r)}_i=0$ for every $i=1,\dots,n$ from (\ref{A7}). Thus, $z^{(r)}=0$. Combining the
above results, we obtain
\begin{equation}\label{A6'}
{\cal Q}_1 + {\cal Q}_2= \sum_{j=1}^{r-1}(y^{(j)}+z^{(j)})^T\theta\ x^{(j)},
\end{equation}
where
\begin{equation}\label{A8}
y^{(j)}_i+z^{(j)}_i=\left(\sum_{k\in\omega_{ij}}x^{(k)}_i\right) + \left(\sum_{k\in\eta_{ij}
}x^{(k)}_i\right) + x^{(j)}_i.
\end{equation}
In the sum (\ref{A8}), every $k\in \omega_{ij}\cap \eta_{ij}$ gives rise to a double appearance of
the term $x^{(k)}_i$ and consequently all such terms vanish. As $\eta_{ij}\subseteq \omega_{ij}$,
we obtain
\begin{equation}\label{A9}y^{(j)}_i+z^{(j)}_i=\left(\sum_{k\in\omega_{ij}\setminus\eta_{ij}}x^{(k)}_i\right) +
x^{(j)}_i.\end{equation} When using (\ref{A9}) to calculate (\ref{A6'}), the terms $x^{(j)}_i$ in
(\ref{A9}) do not contribute to the sum, as they give rise to terms ${x^{(j)}}^T\theta\ x^{(j)}$ in
(\ref{A6'}), which are equal to zero since $\theta$ is symmetric. Thus, defining the vectors
$u^{(j)}$ by
\[u_i^{(j)}:=\left(\sum_{k\in\omega_{ij}\setminus\eta_{ij}}x^{(k)}_i\right),\] (\ref{A6'})
becomes
\begin{equation}\label{Q}
{\cal Q}(X)= \sum_{j=1}^{r-1}{u^{(j)}}^T\theta\ x^{(j)}.
\end{equation}
Note that the set $\omega_{ij}\setminus\eta_{ij}$ consists of all $k\in\{1, \dots, j-1\}$ which
belong to some path $p\neq p(j)$ in ${\cal R}(B_i)$ such that fin$(p)\geq j$. We now show that
whenever $j$ and $l$ belong to the same maximal right path of $B_i$, one has
$\omega_{ij}\setminus\eta_{ij}=\omega_{il}\setminus\eta_{il}$ and consequently
$u_i^{(j)}=u_i^{(l)}$. To see this, fix $i$ and consider arbitrary nodes $j$ and $l$  which lie on
the same maximal right path of $B_i$. Without loss of generality we can assume that $j < l$.
Denoting by $j'$ the right son of $j$, we prove that
$\omega_{ij}\setminus\eta_{ij}=\omega_{ij'}\setminus\eta_{ij'}$: indeed, if $j'=j+1$ then $j$ does
not have a left son (due to the canonical labelling of the nodes) and the assertion follows
trivially; if on the other hand $j'>j+1$ then $j$ has a left subtree. However, the maximal right
paths $p$ in this subtree do not contribute to $\omega_{ij'}\setminus\eta_{ij'}$, as they all
satisfy fin$(p)< j'$ (which is again due to the canonical labelling of the nodes). Therefore
$\omega_{ij}\setminus\eta_{ij}=\omega_{ij'}\setminus\eta_{ij'}$ and iteration of this argument
shows that $\omega_{ij}\setminus\eta_{ij}=\omega_{il}\setminus\eta_{il}$.

We will now use the above property of the $u^{(j)}$'s to show that ${\cal Q}(X)=0$. Let us consider
the first binary tree $B_1$ and suppose that $(1,2,3)$ is a maximal right path of this tree. This
example is chosen for notational convenience, but the argument will work for any maximal right path
in any tree. Thus, we have $u_1^{(1)}=u_1^{(2)}=u_1^{(3)}\equiv u$. Denoting $u^{(j)}=(u, v^{(j)})$
for $j=1,2,3$, the relevant terms in (\ref{Q}) are
\begin{eqnarray}&&{u^{(1)}}^T\theta\ x^{(1)}+{u^{(2)}}^T\theta\ x^{(2)}+{u^{(3)}}^T\theta\
x^{(3)}\nonumber\\&& =\ (u, { v^{(1)}})^T \theta\ x^{(1)}+(u, {v^{(2)}})^T \theta\
x^{(2)}+(u, {v^{(3)}})^T \theta\ x^{(3)}\nonumber\\
&&=\ (u,\ 0\ )^T \theta (x^{(1)}+x^{(2)}+x^{(3)})\nonumber\\&&\ +\ (0, {v^{(1)}})^T \theta\
x^{(1)}+(0, {v^{(2)}})^T \theta\ x^{(2)}+(0, {v^{(3)}})^T \theta\ x^{(3)}\nonumber
\end{eqnarray}
In the r.h.s. of the last equality, the first term is equal to \[u\ \theta_1^T
(x^{(1)}+x^{(2)}+x^{(3)}),\] which is equal to zero from (\ref{recall_VB}), since $(1,2,3)$ is a
maximal right path of $B_1$. Applying this argument to all the maximal right paths of the trees in
${\cal B}$ shows that indeed ${\cal Q}=0$ on the space $V_{\cal B}(G)$. This ends the proof. \hfill
$\square$

\section{Conclusion}

In this paper, we have considered a complete family of local invariants of stabilizer codes and we
have given a translation of these invariants into the binary representation of the stabilizer
formalism. In particular, we have related invariants to dimensions of binary subspaces which depend
only on the generator matrix of a code. The aim of this investigation is mainly to provide a tool
to study the structure of equivalence classes of codes. We note that some important issues in the
present matter remain to be settled: firstly, it is to date not clear how a finite complete set of
invariants can be constructed, i.e., what the minimal degree $r$ is such that the values of the
invariants of degree smaller than $r$ determine the local equivalence class of any stabilizer code.
Secondly, there is the question whether it is sufficient to consider local Clifford operations in
order to recognize local equivalence of stabilizer codes. In other words, are two stabilizer codes
locally equivalent if and only if they are equivalent under the local Clifford group? We believe
that the results in this paper are a significant step towards answering these questions.

\begin{acknowledgments}
MVDN thanks M. Hein, for interesting discussions concerning local equivalence of stabilizer states.
Dr. Bart De Moor is a full professor at the Katholieke Universiteit Leuven, Belgium. Research
supported by Research Council KUL: GOA-Mefisto 666, GOA-Ambiorics, several PhD/postdoc and fellow
grants; Flemish Government: -   FWO: PhD/postdoc grants, projects, G.0240.99 (multilinear algebra),
G.0407.02 (support vector machines), G.0197.02 (power islands), G.0141.03 (Identification and
cryptography), G.0491.03 (control for intensive care glycemia), G.0120.03 (QIT), G.0452.04 (QC),
G.0499.04 (robust SVM), research communities (ICCoS, ANMMM, MLDM); -   AWI: Bil. Int. Collaboration
Hungary/ Poland; -   IWT: PhD Grants, GBOU (McKnow) Belgian Federal Government: Belgian Federal
Science Policy Office: IUAP V-22 (Dynamical Systems and Control: Computation, Identification and
Modelling, 2002-2006), PODO-II (CP/01/40: TMS and Sustainibility); EU: FP5-Quprodis;  ERNSI; Eureka
2063-IMPACT; Eureka 2419-FliTE; Contract Research/agreements: ISMC/IPCOS, Data4s, TML, Elia, LMS,
IPCOS, Mastercard; QUIPROCONE; QUPRODIS.

\end{acknowledgments}

\bibliography{localinvgraph_R4}

\begin{thebibliography}{10}

\bibitem{Gott}
D.~Gottesman.
\newblock {\em Stabilizer codes and quantum error correction}.
\newblock PhD thesis, Caltech, 1997.
\newblock quant-ph/9705052.

\bibitem{codeGF4}
A.R. Calderbank, E.M. Rains, P.W. Shor, and N.J.A. Sloane.
\newblock Quantum error correction via codes over gf(4).
\newblock {\em IEEE transactions on information theory}.
\newblock quant-ph/9608006.

\bibitem{QCQI}
I.~Chuang and M.~Nielsen.
\newblock {\em Quantum computation and quantum information}.
\newblock Cambridge University press, 2000.

\bibitem{pol_inv_suff}
A.L. Onishchik and E.B. Vinberg.
\newblock {\em Lie groups and algebraic groups}.
\newblock springer, Berlin, 1990.

\bibitem{finit_gen}
T.A. Springer.
\newblock {\em Invariant theory}, volume 585 of {\em Lecture notes in
  mathematics}.
\newblock Springer, Berlin, 1977.

\bibitem{invarqubit}
M.~Grassl, M.~R\"otteler, and T.~Beth.
\newblock Computing local invariants of qubit systems.
\newblock {\em Phys.Rev. A}, 58:1833--1839, 1998.
\newblock quant-ph/9712040.

\bibitem{RainsPol}
E.M. Rains.
\newblock Polynomial invariants of quantum codes.
\newblock quant-ph/9704042.

\bibitem{entgraphstate}
M.~Hein, J.~Eisert, and H.J. Briegel.
\newblock Multi-party entanglement in graph states.
\newblock quant-ph/0307130.

\bibitem{localcliffgraph}
M.~Van~den Nest, J.~Dehaene, and B.~De~moor.
\newblock Graphical description of the action of local clifford operations on
  graph states.
\newblock {\em Phys. Rev. A}, 69:022316, 2004.
\newblock quant-ph/0308151.

\bibitem{1wayQC}
R.~Raussendorf, D.E. Browne, and H.J. Briegel.
\newblock Measurement-based quantum computation with cluster states.
\newblock {\em Phys. Rev. A}, 68:022312, 2003.
\newblock quant-ph/0301052.

\bibitem{graphbriegel}
W.~D\"ur, H.~Aschauer, and H.J. Briegel.
\newblock Multiparticle entanglement purification for graph states.
\newblock {\em Phys. Rev. Lett.}, 91:107903, 2003.
\newblock quant-ph/0303087.

\bibitem{stabgraphcode}
D.~Schlingemann.
\newblock Stabilizer codes can be realized as graph codes.
\newblock quant-ph/0111080.

\bibitem{stab_clif_GF2}
J.~Dehaene and B.~De~Moor.
\newblock The clifford group, stabilizer states, and linear and quadratic
  operations over gf(2).
\newblock {\em Phys. Rev. A}, 68:042318, 2003.
\newblock quant-ph/0304125.

\end{thebibliography}

\end{document}